\documentclass[12pt]{iopart}
\pdfoutput=1
\expandafter\let\csname equation*\endcsname\relax
\expandafter\let\csname endequation*\endcsname\relax

\usepackage{color}
\usepackage[breakwords]{truncate}
\usepackage[font={small}]{caption}
\usepackage{graphicx}
\usepackage{cite} 
\usepackage{subcaption}
\usepackage{etoolbox}
\usepackage[section]{placeins}
\usepackage[sectionbib]{chapterbib}
\usepackage{amsfonts}
\usepackage{amsmath}
\usepackage[bookmarks,bookmarksopen,bookmarksdepth=2]{hyperref}
\usepackage{bm}
\makeatletter
\newrobustcmd{\fixappendix}{%
  \patchcmd{\l@section}{1.5em}{7em}{}{}%
  \patchcmd{\l@subsection}{2.3em}{7em}{}{}%
}
\makeatother

\appto\appendix{
\addtocontents{toc}{\fixappendix}
\addtocontents{toc}{\protect\setcounter{tocdepth}{1}}}

\newcommand{\bea}{\begin{eqnarray}}
\newcommand{\eea}{\end{eqnarray}}
\begin{document}
\title[Statistics of the maximum of a Brownian motion in confined geometries]{Statistics of the maximum and the convex hull of a Brownian motion in confined geometries}
\author{Benjamin De Bruyne}
\address{LPTMS, CNRS, Univ.\ Paris-Sud, Universit\'e Paris-Saclay, 91405 Orsay, France}
\author{Olivier Bénichou}
\address{Laboratoire de Physique Th\'eorique de la Mati\`ere Condens\'e,
CNRS/Sorbonne University, 4 Place Jussieu, 75005 Paris, France}
\author{Satya N. Majumdar}
\address{LPTMS, CNRS, Univ.\ Paris-Sud, Universit\'e Paris-Saclay, 91405 Orsay, France}
\author{Grégory Schehr}
\address{Sorbonne Universit\'e, Laboratoire de Physique Th\'eorique et Hautes Energies, CNRS UMR 7589, 4 Place Jussieu, 75252 Paris Cedex 05, France}
\eads{\mailto{benjamin.debruyne@centraliens.net},
\mailto{benichou@lptmc.jussieu.fr},
 \mailto{satya.majumdar@universite-paris-saclay.fr},
\mailto{gregory.schehr@u-psud.fr}}

\begin{abstract}
 We consider a Brownian particle with diffusion coefficient $D$ in a $d$-dimensional ball of radius $R$ with reflecting boundaries. We study the maximum $M_x(t)$ of the trajectory of the particle along the $x$-direction at time $t$. In the long time limit, the maximum converges to the radius of the ball $M_x(t) \to R$ for $t\to \infty$. We investigate how this limit is approached
  and obtain an exact analytical expression for the distribution of the fluctuations $\Delta(t) = [R-M_x(t)]/R$ in the limit of large $t$ in all dimensions. We find that the distribution of $\Delta(t)$ exhibits a rich variety of behaviors depending on the dimension $d$. These results are obtained by establishing a connection between this problem and the narrow escape time problem. We apply our results in $d=2$ to study the convex hull of the trajectory of the particle in a disk of radius $R$ with reflecting boundaries. We find that the mean perimeter $\langle L(t)\rangle$ of the convex hull exhibits a slow convergence towards the perimeter of the circle $2\pi R$ with a stretched exponential decay $2\pi R-\langle L(t)\rangle \propto \sqrt{R}(Dt)^{1/4} \,e^{-2\sqrt{2Dt}/R}$. Finally, we generalise our results to other confining geometries, such as the ellipse with reflecting boundaries. Our results are corroborated by thorough numerical simulations.
\end{abstract}

\section{Introduction and summary of the main results}
\subsection{Introduction}
Brownian motion (BM) provides a natural framework to study the motion of particles in interaction with their environment \cite{einstein06}. In its simplest form, a $d$-dimensional BM $\bm{x}(t)$ with diffusion coefficient $D$ evolves according to the Langevin equation
\begin{align}
  \bm{\dot x}(t) &=\sqrt{2D}\, \bm{\eta}(t)\,,
  \label{eq:eom}
\end{align}
where $\bm{x}(0)=\bm{0}$ is the initial position and where $\bm{\eta}(t)$ is a $d$-dimensional white noise with zero mean $\langle \bm{\eta}(t)\rangle=\bm{0}$ and delta correlated independent components $\langle \eta_i(t)\eta_j(t')\rangle=\delta_{i,j}\delta(t-t')$. BM has proved to describe a wide range of phenomena ranging from chemical reactions \cite{Smol,Krapivsky} to financial stock markets \cite{Bachelier,Bouchaud} and all the way to astronomy \cite{chandrasekhar}. On the theoretical side, BM has attracted a lot of interest both in physics and in mathematics \cite{Feller,Pitman}. One particular active area of research lies in the study of the extreme value statistics (EVS). EVS play a central role in the description of extreme events, such as earthquakes, floods, wildfires, which are rare by nature but can have devastating consequences \cite{Katz02,majumdar2020}. Recently, it has been realised that EVS play also an important role in various systems in random matrix theory \cite{Tracy94,Tracy96,Majumdar14}, fluctuating interfaces \cite{Raychaudhuri01,Gyorgyi03,Majumdar04,Majumdar05c,Gyorgyi07,Burkhardt07,Schehr06,Rambeau09,debruyne21} and computer science problems \cite{Krapivsky00,Majumdar00,Majumdar02,Majumdar05a}. Contrary to the EVS of independently and identically distributed random variables, which are well understood \cite{Gumbel,Fisher,Gnedenko,Leadbetter}, much less is known about the EVS of strongly correlated random variables. Brownian motion being an example of strongly correlated systems, it constitutes a prominent toy model where some analytical progress can be made \cite{majumdar2020,Bray,RednerGuide,Benichou11,Majumdar05,Majumdar10,Aurzada15}.

One of the simplest examples of extreme observables of strongly correlated systems is the maximum $M(t)$ of a one-dimensional Brownian motion $x(t)$ over the time interval $[0,t]$:
\begin{align}
  M(t) = \max_{0\leq\tau\leq t} \left[x(\tau)\right]\,,\label{eq:Min} 
\end{align}
for which the probability distribution $P_M(z,t)$ is given by a half-Gaussian \cite{majumdar2020}
\begin{align}
  P_M(z,t) = \frac{\Theta(z)}{\sqrt{\pi D t}}e^{-\frac{z^2}{4 D t}}\,,\label{eq:Pmz}
\end{align}
where $\Theta(z)$ is the Heaviside step function such that $\Theta(z)=1$ if $z>0$ and $\Theta(z)=0$ otherwise. In particular, from the distribution (\ref{eq:Pmz}), one can easily obtain that the average maximum $M(t)$ of a free Brownian motion grows like
\begin{align}
  \langle M(t) \rangle = \frac{2\sqrt{Dt}}{\sqrt{\pi}}\,.\label{eq:Mavgi}
\end{align}
This observable (\ref{eq:Mavgi}) has found several applications in statistical physics, such as in
the famous Smoluchowski flux problem \cite{Toussaint83,Ziff91,Majumdar06a,Ziff07,Ziff09,Franke12}. Various other extreme observables, such as temporal correlations \cite{Benichou16a,Benichou16b}, extremal times \cite{Mori19a,Mori19b,Mori19c,Schehr10} and other functionals of the maximum \cite{Levy40,Perret13,Perret15} have been obtained for the case of Brownian motion. Extreme observables were also studied for variants of Brownian motion \cite{Majumdar08}, such as for Brownian bridges, excursions and meanders, as well as for non-Markovian processes, such as fractional Brownian motion \cite{Molchan99,Delorme16,Delorme16b,Delorme17,Sadhu18}, the random acceleration process \cite{Burkhardt07,Majumdar10b} and the run-and-tumble particle \cite{Malakar18,Mori20a,Mori20b,Debruyne21b}. EVS have found a particularly nice application in the description of the convex hull of a Brownian motion and other stochastic processes \cite{Kac54,Spitzer56,Snyder,Kabluchko17a,Kabluchko17b,Schawe17,Claussen15,Letac80,Dumonteil13,Reymbaut11,Grebenkov17,Majumdar21,RandonFurling09,Majumdar10c,Schawe18,Chupeau15a,Chupeau15b}, which serves, for instance, as a simple model to describe the extension of the territory of foraging animals in behavioral ecology \cite{Berg93,Bartumeus05,Murphy92,Worton95,Giuggioli11}. This connection has been made possible due a method developed in \cite{Letac80,RandonFurling09,Majumdar10c} which relies on Cauchy's formula \cite{Cauchy32} for convex curves, and states that the average length $\langle L(t)\rangle$ of the convex hull  of a stochastic process in $d=2$ can be expressed in terms of the expected maximum $\langle M(t,\theta)\rangle=\langle \max_t[\mathbf{x}(t) \cdot \mathbf{e}_\theta]\rangle$ in the direction $\mathbf{e}_\theta$ as \cite{RandonFurling09,Majumdar10c}:
\begin{align}
\langle L(t) \rangle &=  \int_0^{2\pi} d\theta \langle M(t,\theta)\rangle\,.\label{eq:LC}
\end{align}
Using the isotropy of BM in $d=2$, one finds directly from (\ref{eq:Mavgi}) and (\ref{eq:LC}) that the mean perimeter of a free BM grows as $\langle L(t) \rangle = 2\pi\langle M(t) \rangle= \sqrt{8\pi D t}$ \cite{Letac80}. Most of the current results on the convex hull of BM concern isotropic processes in an unconfined two-dimensional geometry. Nevertheless, in many practical situations, the process takes place in the presence of boundaries that may limit the growth of the convex hull, such as a river in the context of foraging animals. Recently, the effect of such boundary was studied for a planar Brownian motion in the presence of an infinite reflecting wall. It was shown that the presence of the wall breaks the isotropy of the process and induces a non-trivial effect on the convex hull of the Brownian motion \cite{Chupeau15a,Chupeau15b}. However, the question of the growth of the convex hull of a stochastic process, and more generally of its EVS, in a closed confining geometry remains largely open. The main goal of this paper is to address this question for the case of Brownian motion. 

\begin{figure}[t]
  \begin{center}
    \includegraphics[width=0.4\textwidth]{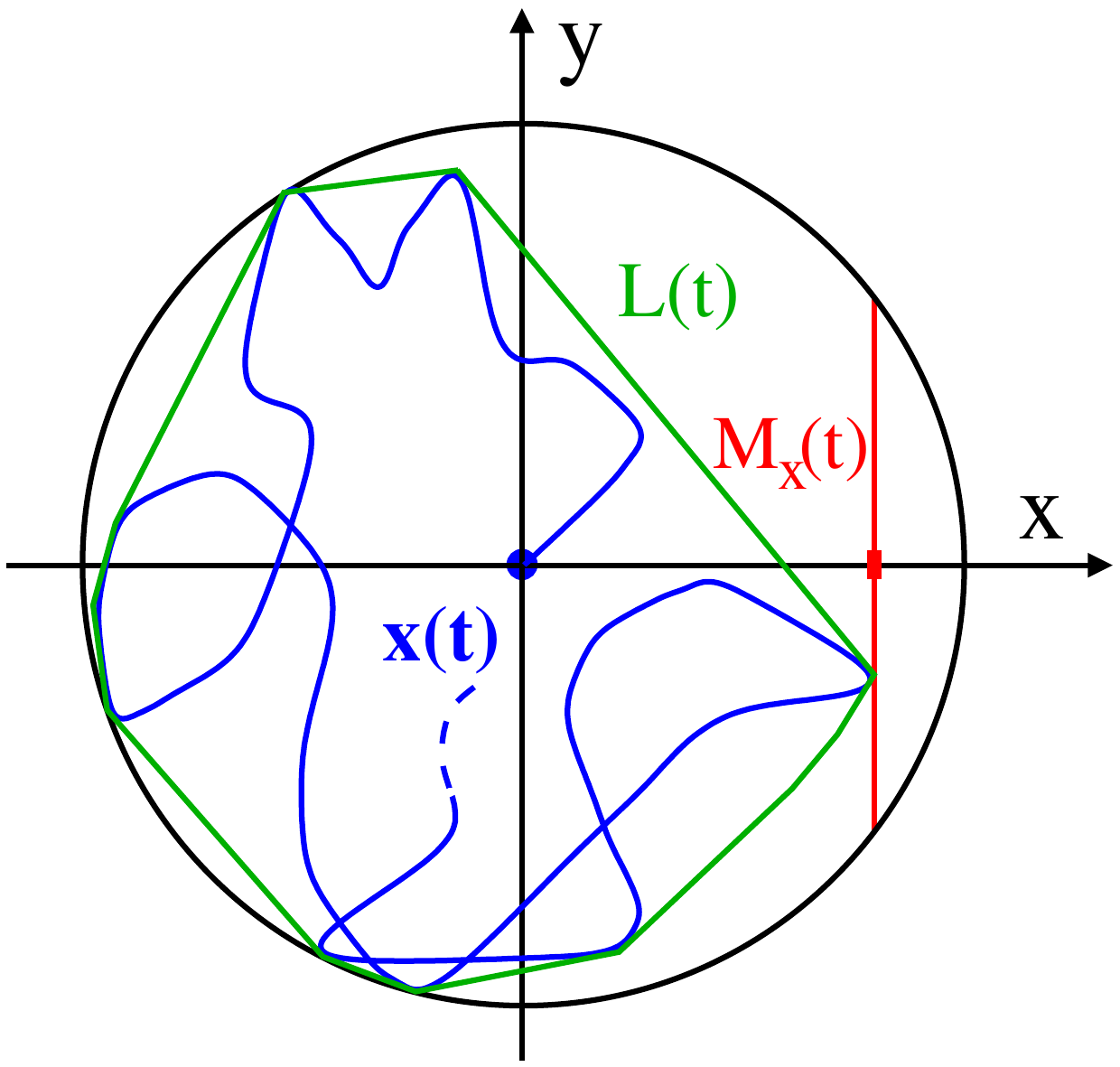}
    \caption{Schematic representation of a planar Brownian motion $\mathbf{x}(t)$ (blue line) in $d=2$ evolving in a disk (black line) of radius $R$ with reflecting boundary conditions. The length of the convex hull $L(t)$ is the length of the convex envelope of the motion up to time $t$ (green line). The maximum in the $x$-direction of the process $M_x(t)=\max_t[\mathbf{x}(t) \cdot \mathbf{e}_x]$ is depicted by the red dash. For $t\to \infty$, the maximum tends to the radius of the circle $M_x(t)\to R$. The distribution of the fluctuations $\Delta(t)=[R-M_x(t)]/R$ follows a non-trivial behavior (see expressions (\ref{eq:dist1i}) for $d=2$ and also (\ref{eq:Mx3di}) for $d\geq 3$). }
    \label{fig:traj}
  \end{center}
\end{figure}

In this paper, we study a Brownian motion (\ref{eq:eom}) confined in a $d$-dimensional ball of radius $R$ with reflecting boundaries. We investigate the growth of the maximum of the process $M_x(t)=\max_t[\mathbf{x}(t) \cdot \mathbf{e}_x]$ in an arbitrary direction (see figure \ref{fig:traj}), which we set to be the $x$-direction without loss of generality due to the rotational symmetry. 
It is clear that in the limit $t\to \infty$, the maximum will tend toward the radius of the ball, namely
 \begin{align}
   M_x(t) \to R\,,\quad t\to \infty\,.\label{eq:MlimR}
 \end{align}
However, for large but finite $t$, the maximum will fluctuate below this limiting value and the fluctuations $\Delta(t)$ can be described by the relative difference between the radius of the ball and the maximum: 
 \begin{align}
   \Delta(t) = \frac{R-M_x(t)}{R}\,.\label{eq:Delta}
 \end{align}
This observable is \textit{a priori} difficult to study for an arbitrary dimension $d$ as it cannot be reduced to a one-dimensional problem, due to the reflecting boundaries of the ball. Nevertheless, by establishing a connection with a similar albeit different problem, which concerns the narrow escape time \cite{Meyer11,Benichou10,Benichou08,Rupprecht15,Singer0,Singer1,Singer2,Singer3}, we obtain exact analytical expressions for the distribution of $\Delta(t)$ in the large $t$ limit. After deriving these results for an arbitrary dimension $d$, we will focus on the special case of $d=2$, where we will use Cauchy's formula to study the growth of the convex hull of a Brownian motion confined in a disk (see figure \ref{fig:traj}). Finally, we will generalise our results to more general geometries.

 \subsection{Summary of the main results}
 It is useful to summarise our main results. Let us first present our results on the growth of the maximum $M_x(t)$ in the direction $x$ for a Brownian motion confined in a $d$-dimensional ball. We find that the decay of the fluctuations (\ref{eq:Delta}) displays a rich behavior depending on the dimension $d$ of the ball. Our main results can be summarised as follows: 
 \paragraph{Finite time convergence with a non-zero probability in $d\!=\!1$.} 
 In the case of $d=1$, the maximum of a Brownian motion in an interval $[-R,R]$, starting at $x=0$, will converge to $R$ in a finite time with a non-zero probability. The distribution of the maximum $P_{M_x}(z,t)$ reaches a stationary state of the form $P_{M_x}(z,t)\sim \delta(z-R)$ for $t\to\infty$. We find that the next-to-leading order correction at time $t$, which describes the convergence to the stationary state, is given by
 \begin{align}
   P_{M_x}(z,t) &\sim \frac{2 \pi  D t}{(z+R)^3}\cos\left(\frac{\pi R}{2(z+R)}\right) \,e^{-\frac{\pi ^2 D t}{4 (z+R)^2}}
   + \delta(z-R)\,\left(1- \frac{2\sqrt{2}}{\pi}  e^{-\frac{\pi ^2 D t}{16R^2}}\right)\,,\quad t\to \infty\,,\label{eq:dist1di}
 \end{align}
 where the Dirac delta term accounts for all the trajectories that have already reached $R$ at time $t$ and the associated weight is the complementary probability for a Brownian motion to survive in an interval of length $2R$ in the limit $t\to \infty$. Despite the finite time convergence of the fluctuations observed in the distribution of the maximum, we obtain from equation (\ref{eq:dist1di}), that the \emph{average} fluctuations decay exponentially with time:
   \begin{align}
  \langle  \Delta(t)\rangle   &\sim \frac{32 \sqrt{2} R^2 }{\pi ^3 D t}\,e^{-\frac{\pi ^2 D t}{16
   R^2}}\,,\quad t\to \infty\,.\label{eq:Delta1avgi}
    \end{align}
 \paragraph{Exponential decay in $d\!=\!2$.} 
   In the case of $d=2$, the maximum $M_x(t)$ along the $x$-direction of a Brownian motion in a disk of radius $R$, starting from the origin, will reach $R$ in an infinite amount of time. We find that the typical fluctuations decay exponentially with time as
   \begin{align}
 \Delta(t) \sim  A_2\,e^{-2\frac{Dt}{R^2} \chi_2}\,,\quad t\to \infty\,,\label{eq:dist1i}
\end{align}
where $\chi_2$ is a random variable of order $O(1)$ whose probability distribution $f_2(\chi_2)$ is given by
\begin{align}
  f_2(\chi_2) = \frac{1}{\chi_2^2}\,e^{-\frac{1}{\chi_2}}\,, \quad \chi_2 \geq 0\,.\label{eq:chi2i}
\end{align}
In equation (\ref{eq:dist1i}), the amplitude $A_2$ is difficult to compute exactly. Below, we give a heuristic argument, leading to $A_2=2\,e^{\frac{1}{4}}$, which is in good agreement with our simulations. Note that the asymptotic relation (\ref{eq:dist1i}) between the random variables $\Delta(t)$ and $\chi_2$ is valid ``in distribution''.  Despite the exponential decay of the fluctuations in (\ref{eq:dist1i}), we find that the \emph{average} fluctuations decay anomalously as a stretched exponential with time, i.e.,
  \begin{align}
   \langle  \Delta(t)\rangle &\sim 2^{1/4}\sqrt{\pi } \,A_2\, \left(\frac{D t}{R^2}\right)^{1/4} \,e^{-2^{3/2} \sqrt{\frac{D t}{R^2}}}\,,\quad t\rightarrow \infty\,.\label{eq:saddleMxi}
\end{align}
The average fluctuations (\ref{eq:saddleMxi}) therefore behave differently from the typical fluctuations (\ref{eq:dist1i}). This originates from the fact that the distribution $f_2(\chi_2)$ has a heavy tail $f_2(\chi_2) \sim \chi_2^{-2}$ for $\chi_2 \to \infty$ such that its first moment does not exist.
 \paragraph{Power law decay in $d\!\geq\!3$.}
 In the case of $d\geq 3$, the maximum $M_x(t)$ along the $x$-direction of a Brownian motion in a $d$-dimensional ball of radius $R$, starting from the origin, will reach $R$ in an infinite amount of time. We find that the typical fluctuations decay algebraically as
\begin{align}
     \Delta(t) \sim A_d  \left(\frac{R^2}{Dt}\right)^{\frac{2}{d-2}}\, \chi_d\,,\quad t\to \infty\,,\label{eq:Mx3di}
\end{align}
where $\chi_d$ is a random variable of order $O(1)$ whose distribution $f_d(\chi_d)$ is given by
\begin{align}
  f_d(\chi_d) = \frac{d-2}{2}\,\frac{e^{-\chi_d^{\frac{d-2}{2}}}}{\chi_d^{\frac{4-d}{2}}}\,,\quad \chi_d \geq 0\,. \label{eq:chidi} 
\end{align}
Here also, the asymptotic relation (\ref{eq:Mx3di}) between the random variables $\Delta(t)$ and $\chi_d$ is valid ``in distribution''. For $d=3$, we have found that the amplitude is given by $A_3= \frac{\pi^2}{18}$ but we did not find an expression for $A_d$ for $d>3$. From the distribution (\ref{eq:Mx3di}), we find that the average fluctuations decay as a power law with time:
\begin{align}
  \langle\Delta(t)\rangle \sim  A_d \,\Gamma\left(\frac{d}{d-2}\right)\left(\frac{R^2}{Dt}\right)^{\frac{2}{d-2}}\,,\quad t\to \infty\,,\label{eq:Mx3ai}
\end{align}
where $\Gamma(z)$ is the gamma function. The case of $d\geq 3$ is therefore quite different to the case of $d\!=\!2$ as the average fluctuations (\ref{eq:Mx3ai}) and the typical ones (\ref{eq:Mx3di}) scale similarly since the first moment of $\chi_d$ is finite for $d\geq 3$.

After deriving these results in arbitrary dimension $d$, we focus on the special case of $d=2$ and make use of Cauchy formula (\ref{eq:LC}) to study the growth of the convex hull of a Brownian motion in a disk (see figure \ref{fig:traj}). We find that the average length of the convex hull $ \langle L(t)\rangle $ approaches slowly the perimeter of the disk $2\pi R$ as a stretched exponential:
\begin{align}
 2\pi R- \langle L(t)\rangle \sim   2^{5/4}\pi^{3/2} \,A_2\, R \left(\frac{D t}{R^2}\right)^{1/4} \,e^{-2^{3/2} \sqrt{\frac{D t}{R^2}}}\,,\quad t\rightarrow \infty\,.\label{eq:saddleLxi}
\end{align}
Finally in Section \ref{sec:gen}, we generalise our results to other geometries, such as the ellipse, for which we also find a similar stretched exponential decay.

The rest of this paper is organised as follows. In Section \ref{sec:narrow}, we recall some results on the narrow escape time which will be our starting point for the next sections. In Section \ref{sec:dist}, we derive the distribution of the fluctuations of the maximum $M_x(t)$ in arbitrary dimensions. In Section \ref{sec:app}, we apply our results to the convex hull of a two-dimensional Brownian motion in a disk. In Section \ref{sec:gen}, we generalise our results to other geometries. Finally, section \ref{sec:ccl} contains our conclusion and perspectives. Some numerical checks and detailed calculations are presented in \ref{app:check} and \ref{app:saddle}.

\section{Narrow escape time}\label{sec:narrow}
\begin{figure}[t]
  \begin{center}
    \includegraphics[width=0.4\textwidth]{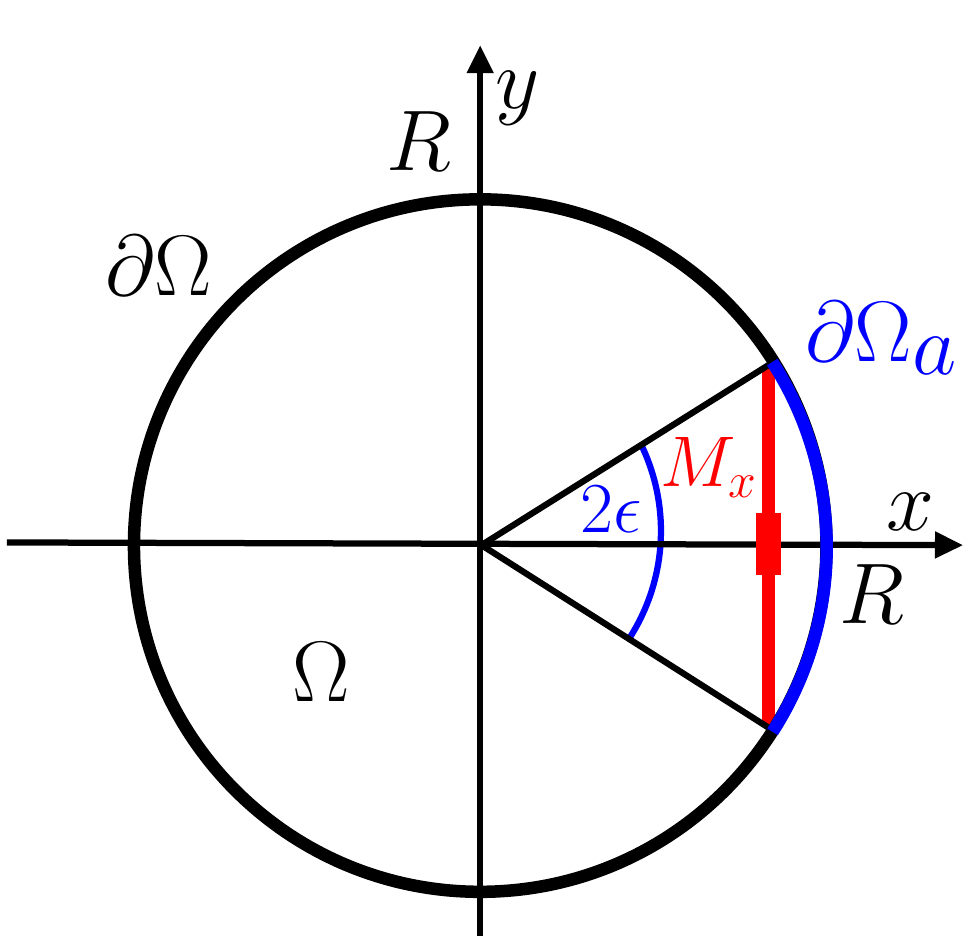}
    \caption{The narrow escape time $\langle T(\epsilon\,|\,\mathbf{x_0})\rangle$ for Brownian motion in a disk $\Omega$ of radius $R$ is the mean first-passage time for the process, starting from $\mathbf{x_0}\in \Omega$, to reach the arc $\partial\Omega_a$ spanned by the angle $2\epsilon$ (blue arc) in the limit of $\epsilon \to 0$, while the complementary boundary $\partial\Omega\backslash \partial\Omega_a$ is reflecting. To study the fluctuations of the maximum in the $x$-direction $M_x$, we assume that when $M_x\to R$, the first-passage time for the process to reach $M_x$ (red line) is asymptotically equal to the first-passage time to reach the arc spanned by the angle $2\epsilon$ with $\epsilon=\arccos\left(\frac{R^2-M_x^2}{R^2}\right)\sim\frac{\sqrt{2(R-M_x)}}{\sqrt{R}}$. }
    \label{fig:circle}
  \end{center}
\end{figure}

In this section, we recall some results on the narrow escape time that will be useful for the next sections. These results are drawn from the recent works on the ``narrow escape problem'' \cite{Meyer11,Benichou10,Benichou08,Rupprecht15,Singer0,Singer1,Singer2,Singer3}. This problem is set as follows. Consider a $d$-dimensional Brownian motion in a closed domain $\Omega$. Let the boundary of the domain $\partial \Omega$ be reflecting everywhere except for a small opening $\partial \Omega_a$, which is absorbing (see figure \ref{fig:circle}).
 The narrow escape problem is then: ``What is the time required for a Brownian motion to escape through the small opening $\partial \Omega_a$?''. This time is known as the narrow escape time (NET) and has received much attention recently due to its importance in various applications such as in biochemical reactions \cite{Singer0,Kolesov07}. To describe the NET, it is convenient to introduce the ratio $\epsilon$ of the size of the opening window over the total size of the boundary:
\begin{align}
  \epsilon =  \frac{|\partial \Omega_a|}{|\partial \Omega|}\ll 1\,.\label{eq:epsilon}
\end{align} 
Clearly, as $\epsilon \to 0$, the mean time to absorption $\langle T(\epsilon\,|\,\mathbf{x_0})\rangle$ starting from $\mathbf{x_0}$ diverges:
\begin{align}
  \langle T(\epsilon\,|\,\mathbf{x_0})\rangle \to \infty\,,\quad \epsilon \to 0\,,\label{eq:Tinf}
\end{align}
when the initial position $\mathbf{x_0}$ of the Brownian walker is located sufficiently far away from the opening window.
As it was shown in a series of papers \cite{Meyer11,Benichou10,Benichou08,Singer0,Singer1,Singer2,Singer3,Rupprecht15}, one can obtain the asymptotic behavior of $\langle T(\epsilon\,|\,\mathbf{x_0})\rangle$ as $\epsilon \to 0$ for a wide range of geometries and in various dimensions. We summarise below the different cases that are relevant for the present work.
\subsection{Mean narrow escape time in $d=2$}
In $d=2$, it was found that for regular domains $\Omega$ that can be conformally mapped to a disk, the NET diverges logarithmically as \cite{Singer0}
\begin{align}
    \langle T(\epsilon\,|\,\mathbf{x_0})\rangle &= \frac{|\Omega|}{D \pi}\left[\ln\left(\frac{1}{\epsilon}\right)+O(1)\right]\,,\label{eq:genT}
\end{align}
where $|\Omega|$ is the size of the domain. In the particular case when $\Omega$ is a disk of radius $R$, it is possible to obtain the next-to-leading order correction in the asymptotic expansion (\ref{eq:genT}). This correction depends on the initial position of the process. When the process starts at the origin of the disk, the NET $ \langle T(\epsilon)\,|\,\mathbf{x_0}=\mathbf{0}\rangle$ is given by \cite{Singer0}
\begin{align}
  \langle T(\epsilon)\,|\,\mathbf{x_0}=\mathbf{0}\rangle = \frac{R^2}{D}\left[\ln\left(\frac{1}{\epsilon}\right)+\ln(2)+\frac{1}{4} + O(\epsilon)\right]\,.\label{eq:Td0}
\end{align} 
On the other hand, the NET averaged over an initial uniform distribution for $\mathbf{x_0}$ in the disk $\overline{\langle T(\epsilon\,|\,\mathbf{x_0})\rangle}$ is given by \cite{Singer0}
\begin{align}
  \overline{\langle T(\epsilon\,|\,\mathbf{x_0})\rangle} =  \frac{R^2}{D}\left[\ln\left(\frac{1}{\epsilon}\right)+\ln(2)+\frac{1}{8} + O(\epsilon)\right]\,.\label{eq:Tda}
\end{align}

\subsection{Mean narrow escape time in $d\geq 3$}
In $d=3$, it was found that for regular bounded domains $\Omega$ with a smooth boundary, the NET through a small disk of radius $\epsilon R$ located on the boundary diverges algebraically as \cite{Singer1}
\begin{align}
    \langle T(\epsilon\,|\,\mathbf{x_0})\rangle = \frac{|\Omega|}{4DR\epsilon} + O[\ln(\epsilon)]\,,\quad \epsilon \to 0\,.\label{eq:Teps3gen}
\end{align}
This result was extended to higher dimensions $d>3$ in \cite{Benichou08} where it was shown that
\begin{align}
    \langle T(\epsilon\,|\,\mathbf{x_0})\rangle \sim  \frac{C_d\,|\Omega|}{DR^{d-2}\,\epsilon^{d-2} }\,,\quad \epsilon\to 0\,,\label{eq:Tepsd}
\end{align}
but the amplitude $C_d$ was not computed. In the following, we will also assume that the next-to-leading correction in (\ref{eq:Tepsd}) grows faster than a constant as it is the case for $d=3$ in (\ref{eq:Teps3gen}).

\subsection{Distribution of the narrow escape time in $d\geq 2$}
In this section, we go beyond the mean value of the NET and we discuss the late time asymptotic of the cumulative distribution $\text{Prob.}\left[ T(\epsilon\,|\,\mathbf{x_0})>t\right]$. In \cite{Meyer11,Benichou10}, it was argued that the cumulative distribution $\text{Prob.}\left[ \tilde T(\epsilon\,|\,\mathbf{x_0})>t\right]$ of the mean first-passage time $\tilde T(\epsilon\,|\,\mathbf{x_0})$ to a small target of size $\epsilon$ located \emph{inside} a domain $\Omega$ behaves in the limit of $t\to \infty$ as 
\begin{align}
  \text{Prob.}\left[ \tilde T(\epsilon\,|\,\mathbf{x_0})>t\right] \sim  \frac{\langle\tilde T(\epsilon\,|\,\mathbf{x_0})\rangle}{ \overline{\langle\tilde T(\epsilon\,|\,\mathbf{x_0})\rangle}}\,\exp\left(-\frac{t}{\overline{\langle\tilde T(\epsilon\,|\,\mathbf{x_0})\rangle}}\right)\,, \quad \epsilon \to 0\,,\quad t \rightarrow\infty\,,\label{eq:PepsT}
\end{align}
where $\langle\tilde T(\epsilon\,|\,\mathbf{x_0})\rangle$ is the mean first-passage time to the target from the initial position $\mathbf{x_0}$ and $\overline{\langle \tilde T(\epsilon\,|\,\mathbf{x_0})\rangle}$ is the same quantity but averaged over an initial uniform position in the domain $\Omega$. It is natural to extend this result to the CDF of the NET which can be considered as a first-passage time to a target located on the boundary of the domain, as it was done in \cite{Rupprecht15} for the case of a spherical domain. In the following of this work, we assume that the asymptotic behavior (\ref{eq:PepsT}) is also valid for the CDF of the NET $\text{Prob.}\left[ T(\epsilon\,|\,\mathbf{x_0})>t\right]$ (see \ref{app:check} for a numerical check).

\section{Distribution of the maximum}\label{sec:dist}
In this section, we derive the distribution of the fluctuations of the maximum $M_x(t)$ in a ball of dimension $d$ with reflecting boundaries. As the fluctuations display a different behavior depending on the dimension $d$, we distinguish the three different cases: $d=1$, $d=2$ and $d\geq 3$. While the case of $d=1$ can be solved straightforwardly, the cases of $d=2$ and $d\geq 3$ are more difficult to study and our approach relies on the results on the NET discussed in the previous section.
\subsection{One-dimensional interval ($d=1$)}
\begin{figure}[t]
  \centering\subfloat[]{%
\includegraphics[width=0.45\textwidth]{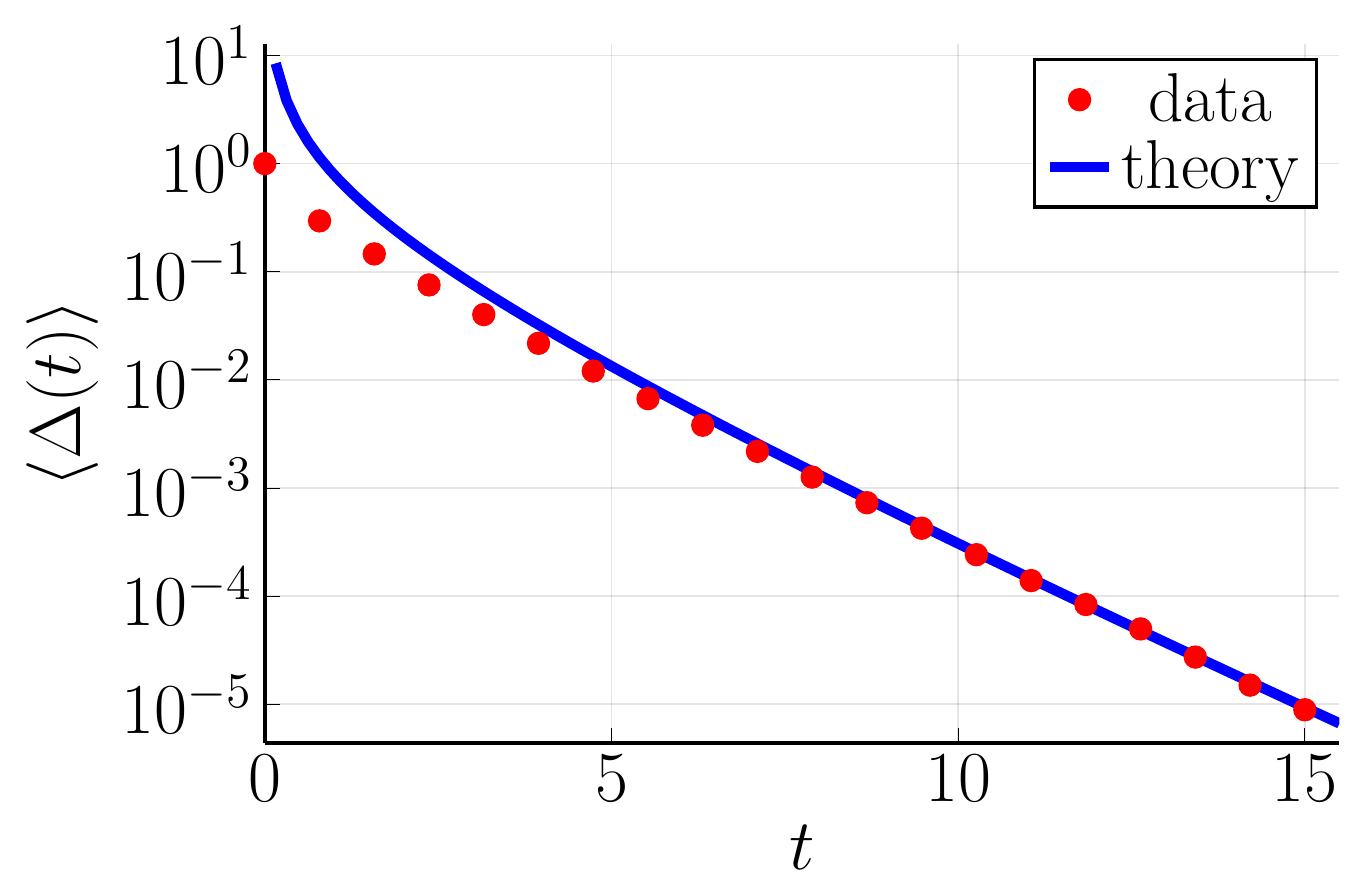}%
}\hfill
\subfloat[]{%
\includegraphics[width=0.45\textwidth]{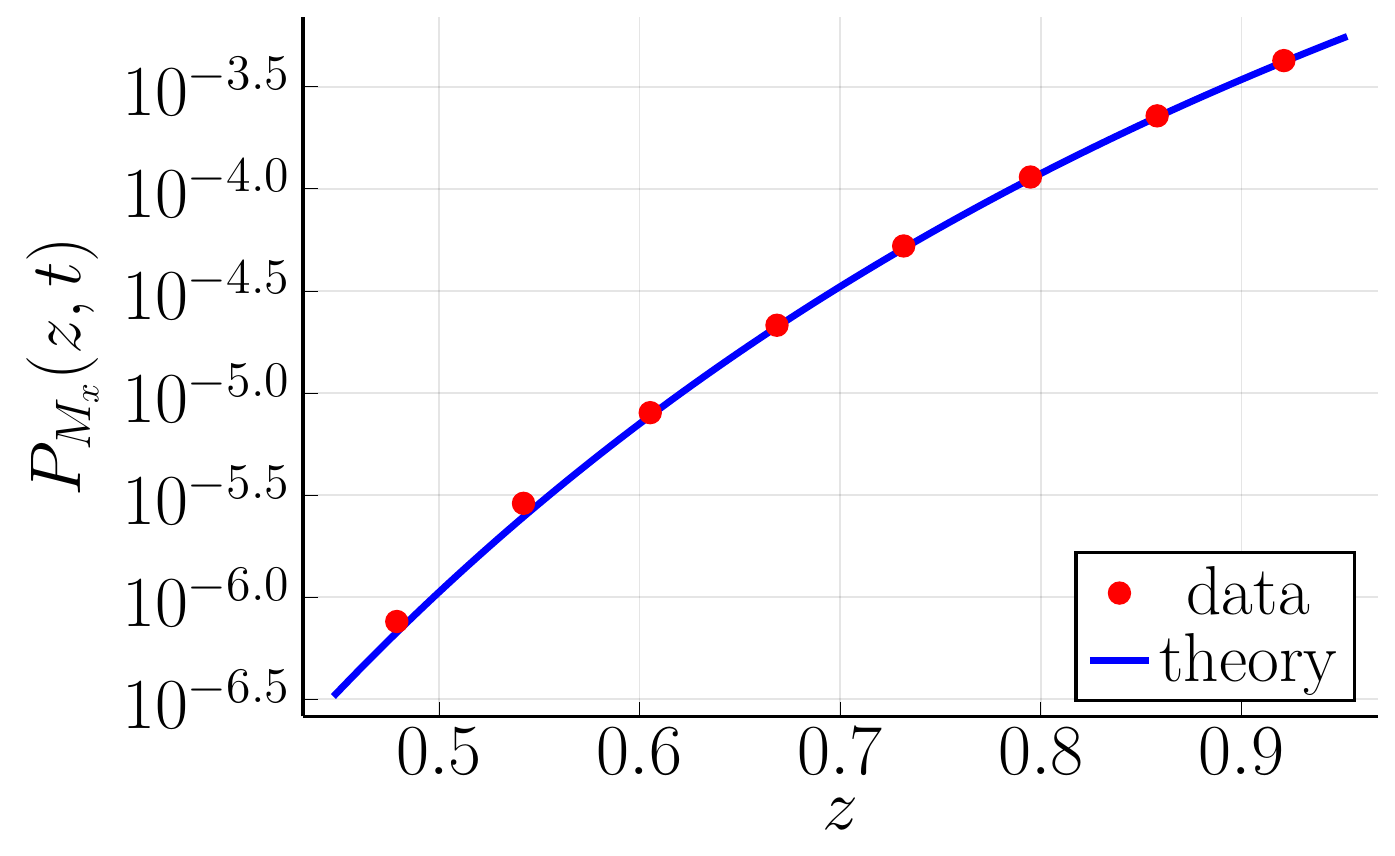}%
}\hfill
    \caption{\textbf{Left panel:} Evolution of the mean fluctuations $\langle \Delta(t )\rangle=\left\langle \frac{R-M_x(t)}{R}\right\rangle$ of a $d\!=\!1$ Brownian motion in an interval $[-R,R]$ with reflecting boundaries. The numerical data (red dots) have been obtained by averaging over $10^6$ trajectories with $dt=10^{-2}$ and are compared to the theoretical prediction (\ref{eq:Delta1avgi}) with $D=1$ and $R=1$.  \textbf{Right panel:} Distribution of the maximum $P_{M_x}(z,t)$ computed numerically (red dots), averaged over $10^9$ realisations with $dt=10^{-4}$ and compared to the theoretical prediction (blue line), given in (\ref{eq:dist1di}) for $D=1$, $R=1$ and $t=15$. The Dirac delta in (\ref{eq:dist1di}) is not shown to fit in the limited window size.  }
    \label{fig:fchi1d}
  \end{figure}

In this section, we consider a one-dimensional Brownian motion in an interval $[-R,R]$ with reflecting boundaries. We assume that the particle starts initially at the origin $x_0=0$. We study the evolution of the maximum $M_x(t)$ as a function of time. The cumulative distribution $\text{Prob.}\left(M_x(t)<M_x\right)$ can be obtained by noting that it is equal to the probability that the diffusive particle did not reach $M_x$ up to time $t$ given that it started from the origin:
\begin{align}
  \text{Prob.}\left(M_x(t)<M_x\right) = \text{Prob.}\left(T_{M_x}>t\right)\,,\label{eq:id1d}
\end{align}
where $T_{M_x}$ is the first-passage time to $M_x$ starting from the origin. The term in the right-hand side in (\ref{eq:id1d}) is simply the survival probability up to time $t$ of a diffusive particle in the interval $[-R,M_x]$, with a reflecting boundary condition at $-R$ and an absorbing one at $M_x$, which in Laplace domain reads \cite{RednerGuide}
\begin{align}
  \int_0^\infty dt\,e^{-st}\, \text{Prob.}\left(M_x(t)<M_x\right)  = \frac{1}{s}\left(1-\frac{\cosh\left(\sqrt{\frac{s}{D}}R\right)}{\cosh\left[\sqrt{\frac{s}{D}}(M_x+R)\right)}\right)\,.\label{eq:FMx1}
\end{align}
In the complex $s$-plane, the first imaginary pole of the Laplace transform (\ref{eq:FMx1}) is located at $s^*$ such that $i \sqrt{\frac{s^*}{D}}(M_x+R)=\pi/2$. By computing the residue at $s^*$, we find that the decay of the cumulative distribution behaves, to leading order, for $t\to\infty$ as
\begin{align}
\text{Prob.}\left(M_x(t)\leq M_x\right) \sim \left\{\begin{array}{ll}\frac{4}{\pi}\cos\left(\frac{\pi R}{2(M_x+R)}\right) \,e^{-\frac{\pi ^2 D t}{4
   (M_x+R)^2}} \,,& M_x<R\,,\\
 1 \,,\,& M_x=R\,,
 \end{array}\right.\quad t \rightarrow\infty\,.
 \label{eq:Fmt2}
\end{align}
Note that the CDF (\ref{eq:Fmt2}) is discontinuous at $M_x\!=\!R$ as $\text{Prob}.\left(M_x(t)<R\right)\neq \text{Prob}.\left(M_x(t)\leq R\right)$. By taking a derivative with respect to $M_x$ of the cumulative distribution (\ref{eq:Fmt2}), we obtain the distribution (\ref{eq:dist1di}) displayed in the introduction. 
This result is in good agreement with numerical data (see right panel in figure \ref{fig:fchi1d}).  The numerical data were obtained by discretising the equation of motion (\ref{eq:eom}) over small time increments $dt$ and by drawing Gaussian random numbers with standard deviation $dx=\sqrt{2Ddt}$ at each time step. The reflecting boundary is implemented by assuming ballistic evolution with elastic reflections within each time steps $dt$. From the distribution (\ref{eq:dist1di}), one can obtain the expected value of the fluctuations $\Delta(t)=[R-M_x(t)]/R$ which is given in (\ref{eq:Delta1avgi}) and is in good agreement with numerical data (see left panel in figure \ref{fig:fchi1d}).

\subsection{Two-dimensional disk ($d=2$)}\label{sec:disk}
\begin{figure}[t]
  \centering\subfloat[]{%
\includegraphics[width=0.45\textwidth]{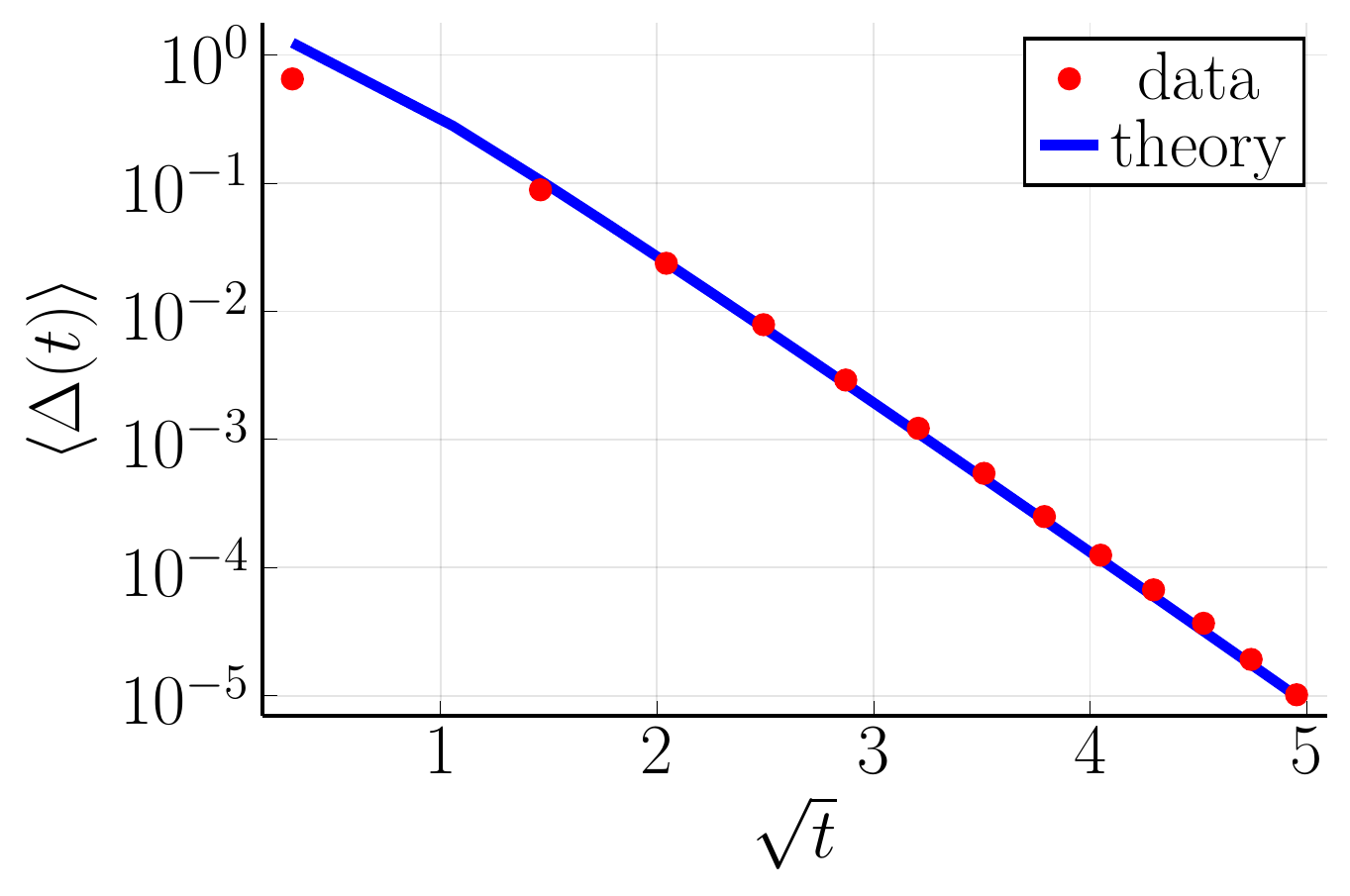}%
}\hfill
\subfloat[]{%
\includegraphics[width=0.45\textwidth]{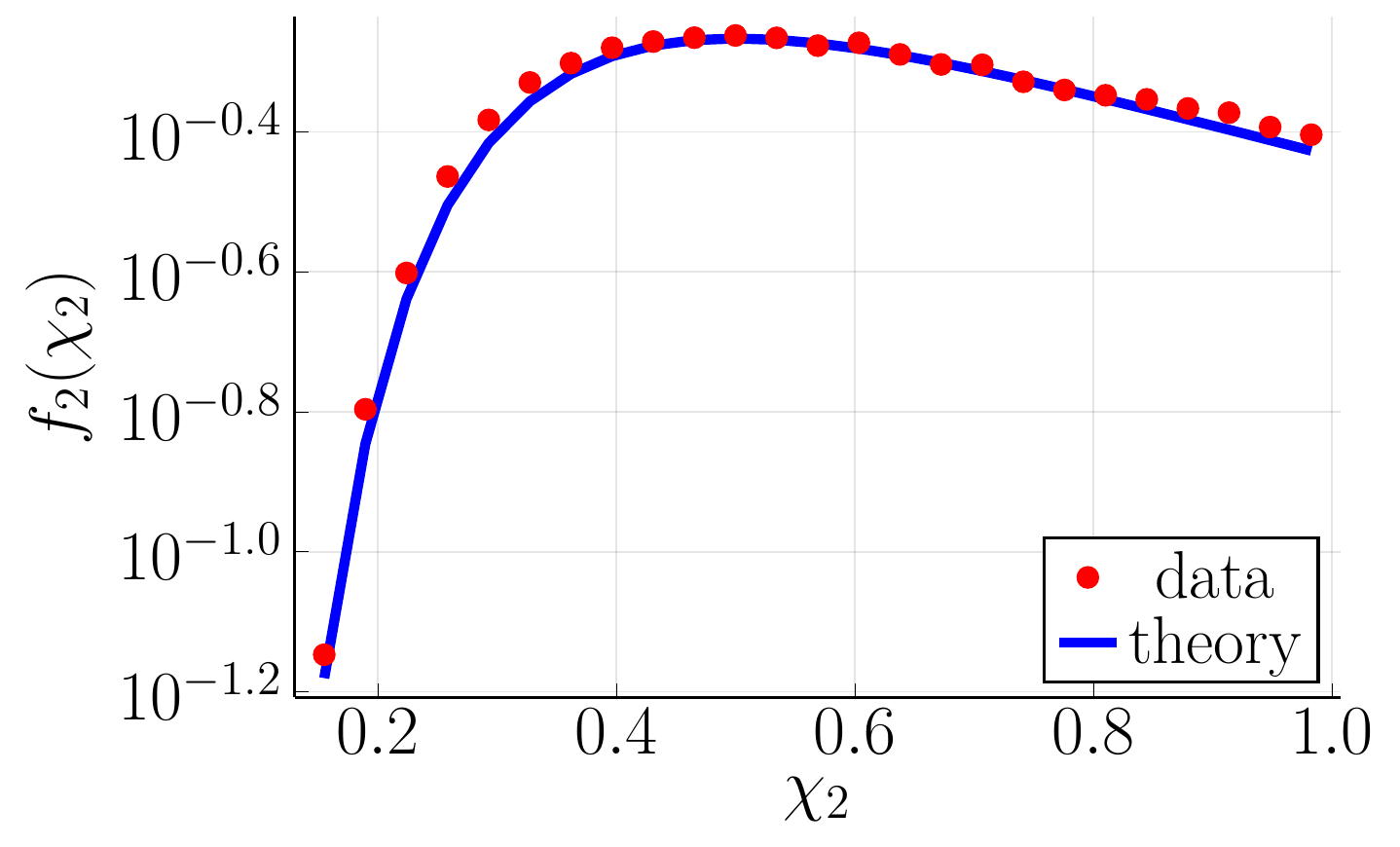}%
}\hfill
    \caption{\textbf{Left panel:} Evolution of the mean fluctuations $\langle \Delta(t )\rangle=\left\langle \frac{R-M_x(t)}{R}\right\rangle$ of a $d\!=\!2$ Brownian motion in a disk of radius $R$ with reflecting boundaries. The numerical data (red dots) have been obtained by averaging over $10^5$ trajectories with $dt=10^{-6}$ and are compared to the theoretical prediction (\ref{eq:saddleMxi}) with $D=1$, $R=1$ and $A_2=2\,e^{\frac{1}{4}}$.  \textbf{Right panel:} Distribution of the rescaled fluctuations $\chi_2=\frac{R^2}{2Dt}\left[\frac{1}{4}-\ln\left(\frac{\Delta(t)}{2R}\right)\right]$ computed numerically (red dots), averaged over $ 10^6$ realisations with $dt=10^{-9}$ and compared to the theoretical prediction (blue line), given in (\ref{eq:chi2i}) with $A_2=2\,e^{\frac{1}{4}}$ for $D=1$, $R=1$ and $t=8$. The plot is limited to $\chi_2<1$ where the discretisation step is much smaller than the difference $R-M_x$. }
    \label{fig:fchi2d}
  \end{figure}

To obtain the distribution of $M_x(t)$ for the case of a two-dimensional Brownian motion in a disk of radius $R$ with reflecting boundaries starting from the origin, we use the NET results from Section \ref{sec:narrow} and rely on the following assumption
\begin{align}
  \text{Prob.}\left(T_{M_x}>t\right) & \sim \text{Prob.}\left[T\left(\epsilon=\sqrt{\frac{2(R-M_x)}{R}}\,\bigg|\,\mathbf{0}\right)>t\right]\,,\quad t\to \infty\,,\label{eq:id}
\end{align}
where $T_{M_x}$ is the first-passage time to $M_x$ starting from the origin and $T(\epsilon|\mathbf{x_0})$ is the NET to reach an arc of angle $2\epsilon$ in a circular domain, given that the Brownian motion started at $\mathbf{x_0}$. In other words, the assumption (\ref{eq:id}) can be stated as follows: the probability to hit the arc of angle $2\epsilon$ for the first time is asymptotically equal to the probability to hit the chord subtended by this angle for the first time in the limit of $\epsilon\to 0$ (see figure \ref{fig:circle}). The assumption (\ref{eq:id}) is an approximation for any finite $t$ but we expect it to be asymptotically exact in the limit $t\to \infty$. Under this assumption, we can further use the identity (\ref{eq:id1d}) and the CDF of the NET (\ref{eq:PepsT}) to obtain that
\begin{align}
  \text{Prob.}\left(M_x(t)<M_x\right) \sim \frac{\left\langle T\left(\epsilon\,|\,\mathbf{0}\right)\right\rangle}{\overline{\left\langle T\left(\epsilon\,|\,\mathbf{x_0}\right)\right\rangle}}\exp\left(-\frac{t}{\overline{\left\langle T\left(\epsilon\,\big|\,\mathbf{x_0}\right)\right\rangle}}\right)\Bigg\rvert_{\epsilon=\sqrt{\frac{2(R-M_x)}{R}}} \,,\quad t \rightarrow\infty\,, \label{eq:Fmt}
\end{align}
where $\overline{\langle T(\epsilon|\mathbf{x_0})\rangle}$ is the NET averaged over an initial uniform position in the disk. We now recall the expressions of the NET for a disk given in (\ref{eq:Td0})-(\ref{eq:Tda}) evaluated at $\epsilon=\frac{\sqrt{2(R-M_x)}}{\sqrt{R}}$:
\begin{subequations}
\begin{align}
  \left\langle T\left(\epsilon\,|\,\mathbf{0}\right)\right\rangle\bigg\rvert_{\epsilon=\sqrt{\frac{2(R-M_x)}{R}}}&\sim \frac{R^2}{2D} \ln\left(\frac{1}{R-M_x}\right)+O(1)\,,\quad M_x\rightarrow R\,,\label{eq:Tapprox1}\\[1em]
  \overline{\left\langle T\left(\epsilon\,|\,\mathbf{x_0}\right)\right\rangle}\bigg\rvert_{\epsilon=\sqrt{\frac{2(R-M_x)}{R}}} &\sim \frac{R^2}{2D} \ln\left(\frac{1}{R-M_x}\right)+O(1)\,,\quad M_x\rightarrow R\,.\label{eq:Tapprox2}
\end{align}
\label{eq:Tapprox}
\end{subequations}
Inserting the NETs (\ref{eq:Tapprox}) in the cumulative distribution (\ref{eq:Fmt}), we find
\begin{align}
  \text{Prob.}\left(M_x(t)<M_x\right) \sim \exp\left(-\frac{2Dt}{R^2\,\ln\left(\frac{1}{R-M_x}\right)+O(1)}\right)\,,\quad t\to \infty\,.\label{eq:Probf}
\end{align}
By taking a derivative with respect to $M_x$ of the cumulative distribution (\ref{eq:Probf}) and by denoting $\chi_2=[R^2\,\ln\left(\frac{1}{R-M_x}\right)+O(1)]/(2Dt)$, one obtains the distribution (\ref{eq:dist1i}) displayed in the introduction with an unknown amplitude $A_2$. In principle, this amplitude can be obtained from the order O(1) term in (\ref{eq:Tapprox2}), which is, unfortunately, not known. If we assume that this next-to-leading order term is the same as the one in the expansion of the NET given in (\ref{eq:Td0})-(\ref{eq:Tda}), we obtain that $A_2=2\,e^{\frac{1}{4}}$. However, it is not clear that we are allowed to do so as the first-passage time to reach the arc of angle $2\epsilon$ and the first-passage time to reach the chord subtended by this angle might differ by finite-size corrections in the limit of $\epsilon\to 0$ (see figure \ref{fig:circle}). Nevertheless, this result is in good agreement with numerical data (see right panel in figure \ref{fig:fchi2d}). From the distribution (\ref{eq:dist1i}), one can compute the average value of the fluctuations which gives
\begin{align}
   \langle \Delta(t )\rangle &=  A_2\,\int_0^\infty \frac{d\chi_2}{\chi_2^2}\,\exp\left[-\left(\frac{2Dt}{R^2} \,\chi_2+\frac{1}{\chi_2}\right)\right]\,.\label{eq:avgDelta2}
\end{align}
   This integral can be computed in the limit $t\to \infty$ by the saddle point method (see \ref{app:saddle}) and we recover the expression (\ref{eq:saddleMxi}) displayed in the introduction.
This result is in good agreement with numerical simulations (see left panel in figure \ref{fig:fchi2d}).
\subsection{$d$-dimensional ball ($d\geq 3$)}
\begin{figure}[t]
  \centering\subfloat[]{%
\includegraphics[width=0.45\textwidth]{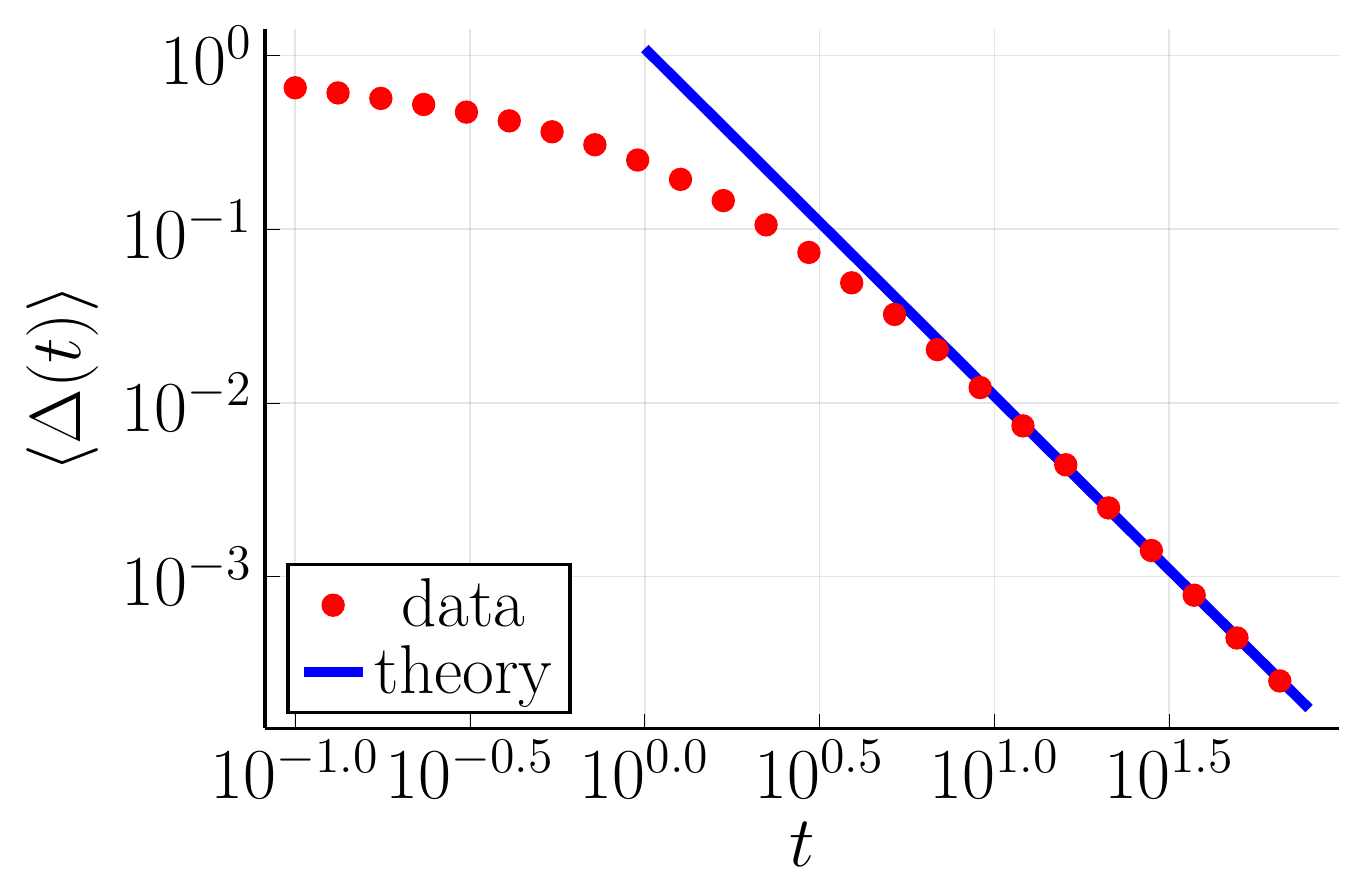}%
}\hfill
\subfloat[]{%
\includegraphics[width=0.45\textwidth]{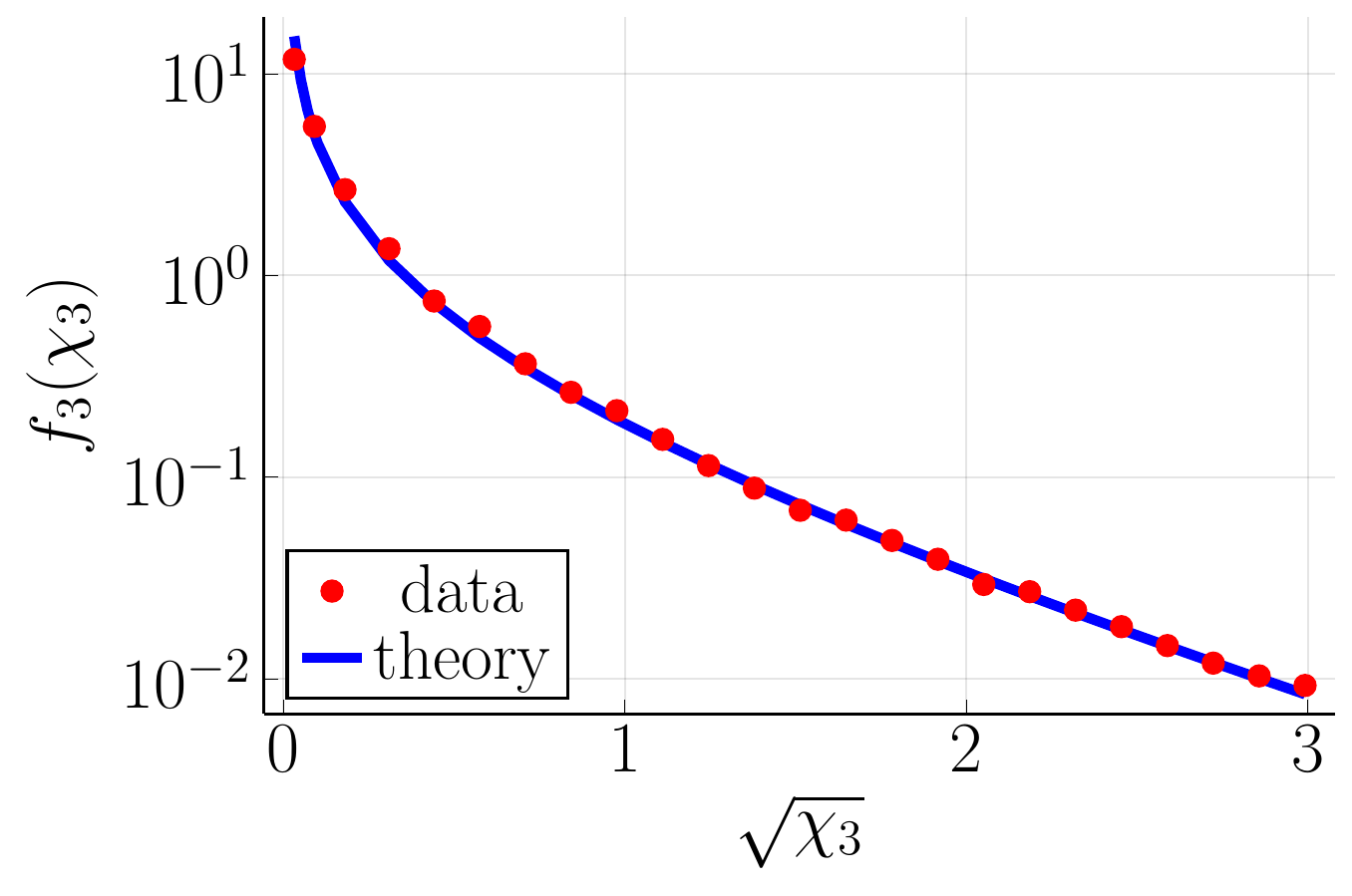}%
}\hfill
    \caption{\textbf{Left panel:} Evolution of the mean fluctuations $\langle \Delta(t )\rangle=\left\langle \frac{R-M_x(t)}{R}\right\rangle$ of a $d=3$ Brownian motion in a sphere of radius $R$ with reflecting boundaries. The numerical data (red dots) have been obtained by averaging over $10^5$ trajectories with $dt=10^{-6}$ and are compared to the theoretical prediction (\ref{eq:Mx3ai}) with $d=3$, $D=1$, $R=1$ and $A_3=\frac{\pi^2}{18}$.  \textbf{Right panel:} Distribution of the rescaled fluctuations $\chi_3 = \frac{18D^2 t^2}{\pi^2 R^5}\Delta(t)$ computed numerically (red dots), averaged over $10^4$ realisations with $dt=10^{-6}$ and compared to the theoretical prediction (blue line), given in (\ref{eq:chidi}) with $A_3=\frac{\pi^2}{18}$, $D=1$, $R=1$, $t=24$ and $d=3$. }
    \label{fig:fchi3d}
  \end{figure}

In this section, we consider a $d$-dimensional Brownian motion in a ball of radius $R$ with reflecting boundaries. We study the evolution of the expected maximum of the $x$-component $M_x(t)$ as a function of time. As in the previous section, we use the NET results from Section \ref{sec:narrow} and rely on the assumption (\ref{eq:id}). The CDF of $M_x(t)$ is therefore given by (\ref{eq:Fmt}). We now recall the expressions of the NET for a $d$-dimensional domain given in (\ref{eq:Tepsd}) evaluated for a ball of radius $R$:
\begin{subequations}
\begin{align}
  \langle T(\epsilon\,|\,\mathbf{0})\rangle & \sim \frac{\tilde C_d\,R^2}{D\,\epsilon^{d-2}} \,,\quad \epsilon \to 0\,,\\
  \overline{\langle T(\epsilon\,|\,\mathbf{x_0})\rangle} &\sim  \frac{\tilde C_d\,R^2}{D\,\epsilon^{d-2}} \,,\quad \epsilon \to 0\,,
\end{align}
 \label{eq:Teps3}
\end{subequations}
where $\tilde C_d$ is an amplitude, which for $d=3$ is given by $\tilde C_3=\frac{\pi}{3}$, and the next-to-leading term is assumed to grow faster than a constant. Inserting the NET (\ref{eq:Teps3}) with $\epsilon=\sqrt{\frac{2(R-M_x)}{R}}$ in the CDF (\ref{eq:Fmt}), we find 
\begin{align}
  \text{Prob.}\left(M_x(t)<M_x\right) \sim \exp\left(-\frac{2^{\frac{d-2}{2}} \,Dt\,(R-M_x)^{\frac{d-2}{2}}}{\tilde C_d\, R^{\frac{d+2}{2}}}\right)\,,\quad t\to \infty\,.\label{eq:FMx3d}
\end{align}
By taking a derivative with respect to $M_x$ of the cumulative distribution (\ref{eq:FMx3d}) and by denoting $\chi_d=[2(Dt)^{\frac{2}{d-2}}(R-M_x)]/(\tilde C_d^{\frac{2}{d-2}}R^{\frac{d+2}{d-2}})$, one obtains the distribution (\ref{eq:Mx3di}) displayed in the introduction. This result is in good agreement with numerical data for $d=3$ (see right panel in figure \ref{fig:fchi3d}). From the distribution (\ref{eq:Mx3di}), one can compute the average value of the fluctuations by using the following identity
\begin{align}
  \int_0^\infty d\chi_d \,\chi_d\,\frac{d-2}{2}\,\frac{e^{-\chi_d^{\frac{d-2}{2}}}}{\chi_d^{\frac{4-d}{2}}} = \Gamma\left(\frac{d}{d-2}\right)\,,\quad d\geq 3\,.\label{eq:idav}
\end{align}
Using the identity (\ref{eq:idav}), we obtain the average value (\ref{eq:Mx3ai}) displayed in the introduction.
This result is in good agreement with numerical data for $d=3$ (see left panel in figure \ref{fig:fchi3d}).
\section{Application to the convex hull in a two-dimensional disk} \label{sec:app}
In this section, we apply our previous results in $d=2$ to study the convex hull of a Brownian motion in a disk of radius $R$ with reflecting boundaries (see figure \ref{fig:traj}). As the Brownian motion explores the disk, its convex hull gradually grows and eventually covers the whole disk. We are interested in the behavior of the mean perimeter $\langle L(t)\rangle$ of the convex hull as a function of time. Due to the confining disk, the mean perimeter is limited in $0\leq \langle L(t)\rangle \leq 2\pi R$. Eventually, the convex hull will cover the whole disk and therefore, we have 
\begin{align}
  \langle L(t)\rangle \sim 2\pi R\,, \quad t \rightarrow \infty\,. \label{eq:Linf}
\end{align}
We wish to determine the speed at which this convergence takes place, i.e.~we would like to find the second order term in the asymptotic expansion (\ref{eq:Linf}).
Due to the isotropy of Brownian motion, the mean perimeter $\langle L(t)\rangle$ is related to the mean maximum  of the process in an arbitrary direction, which we choose to be the $x$-direction $\langle M_x(t)\rangle $, through Cauchy's formula (\ref{eq:LC}):
\begin{align}
  \langle L(t)\rangle =  \int_0^{2\pi} d\theta \langle M(t,\theta)\rangle =2\pi  \langle M_x(t )\rangle\,,\label{eq:Cauchy}
\end{align}
where we used that the motion is isotropic.
Finally, using Cauchy's formula (\ref{eq:Cauchy}) and the asymptotic behavior of the maximum (\ref{eq:saddleMxi}), we find that the asymptotic behavior of the mean perimeter $\langle L(t)\rangle$ of the convex hull is given by:
   \begin{align}
  2\pi R -  \langle L(t )\rangle &\sim   2^{5/4}\pi^{3/2} \,A_2\, R \left(\frac{D t}{R^2}\right)^{1/4} \,e^{-2^{3/2} \sqrt{\frac{D t}{R^2}}}\,,\quad t\rightarrow \infty\,.\label{eq:Linf2}
\end{align}
The result (\ref{eq:Linf2}) is interesting as it indicates a slow convergence, with a stretched exponential, of the mean perimeter of the convex hull $\langle L(t)\rangle$ to the perimeter of the disk $2\pi R$.

\section{Generalisation to other geometries in two dimensions}\label{sec:gen}
In this section, we generalise our results, valid for circular geometries, to the case of other two-dimensional bounded domains. By proceeding similarly to the case of the disk in Section \ref{sec:disk} and by using the general result (\ref{eq:genT}) on the NET for regular domains $\Omega$ that can be mapped conformally to a disk, we find that the fluctuations $\Delta(t)$ of the maximum close to its maximal value in an arbitrary direction behave as
\begin{align}
   \Delta(t) \sim \mathcal{A}_2\,e^{-\frac{\pi Dt}{|\Omega|} \chi_2}\,,\quad t\to \infty\,,\label{eq:flug}
\end{align}
where $\mathcal{A}_2$ is an unknown amplitude which depends on the geometry but is independent of the direction, $|\Omega|$ is the size of the domain and $\chi_2$ is the random variable whose distribution is given in (\ref{eq:chi2i}). In particular, the average fluctuations of the maximum close to its maximal value in an arbitrary direction behave as
\begin{align}
\langle  \Delta(t)\rangle &\sim 2^{1/4}\sqrt{\pi } \,\mathcal{A}_2\, \left(\frac{\pi D t}{|\Omega|}\right)^{1/4} \,e^{-2^{3/2} \sqrt{\frac{\pi D t}{|\Omega|}}}\,,\quad t\rightarrow \infty\,.\label{eq:saddleMg}
\end{align}
In the remaining of this section, we apply this result to study the growth of the mean perimeter of the convex hull of a Brownian motion in an ellipse with reflecting boundaries.
\subsection{Growth of the mean perimeter in an ellipse}
\begin{figure}[ht]
  \begin{center}
    \includegraphics[width=0.4\textwidth]{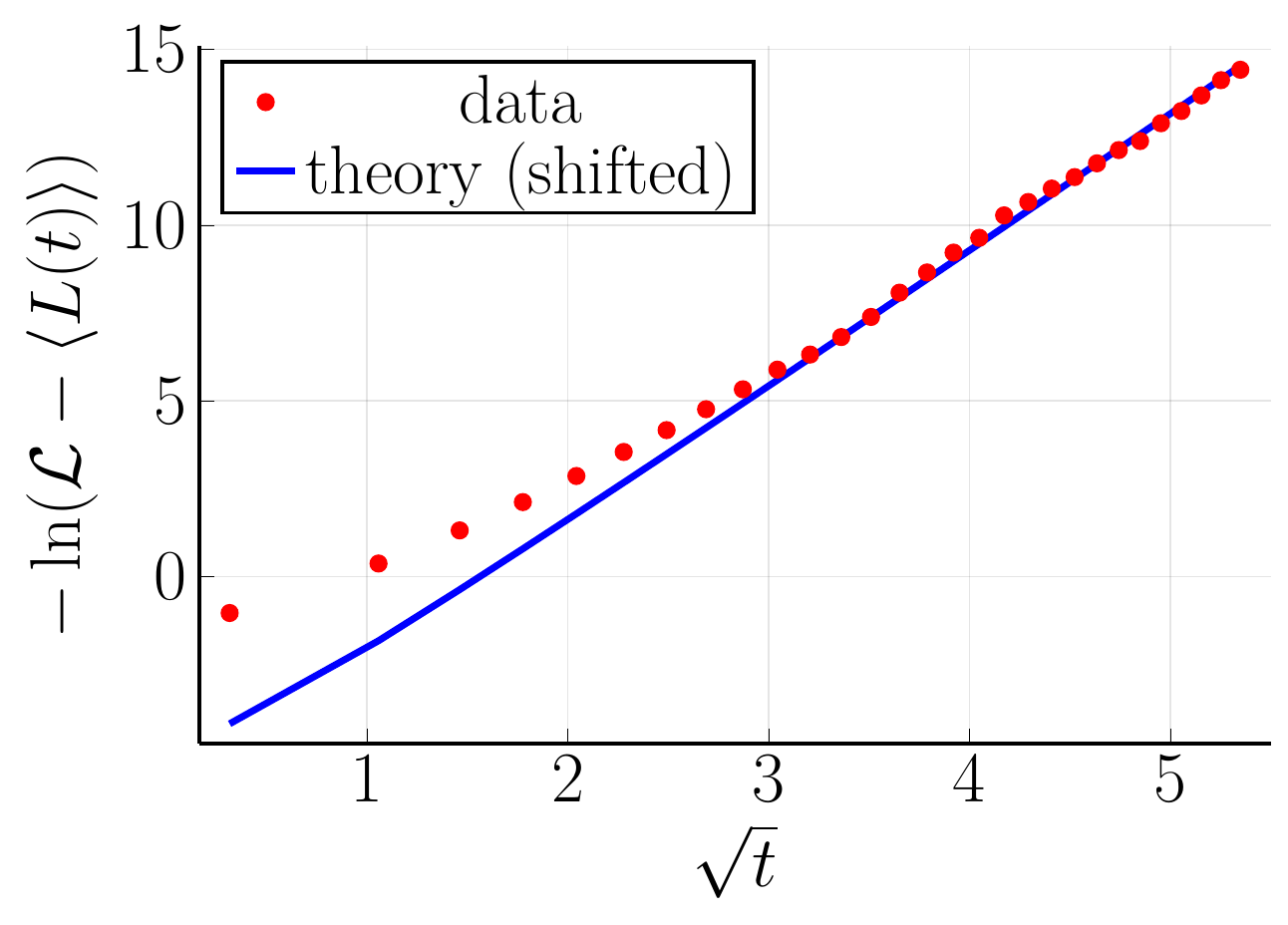}
    \caption{ Evolution of the mean perimeter $\langle L(t)\rangle$ of the convex hull of a $d=2$ Brownian motion in a confining ellipse of semi-axes $a=1$ and $b=1/2$ with reflecting boundaries. The numerical data have been computed by measuring the mean perimeter of the convex hull $\langle L(t)\rangle$ using Graham scan algorithm \cite{Graham72}. The average has been taken over $10^5$ trajectories with $dt=10^{-6}$ and $D=1$. The theoretical curve (\ref{eq:Lell}) has been shifted for better comparison (since we do not predict the exact amplitude). }
    \label{fig:decay}
  \end{center}
\end{figure}

In this section, we consider a two-dimensional Brownian motion in a reflecting elliptical geometry defined by
\begin{align}
  \frac{x^2}{a^2} + \frac{y^2}{b^2} =1\,,\label{eq:ellipse}
\end{align}
 where $a\geq b>0$ are the ellipse semi-axes. It is well-known that the area of such geometry is given by $|\Omega|=\pi ab$. Therefore, the average fluctuations (\ref{eq:saddleMg}) of the maximum close to its maximal value in an arbitrary direction are given by
 \begin{align}
   \langle  \Delta(t)\rangle &\sim 2^{1/4}\sqrt{\pi } \,\mathcal{A}_2\, \left(\frac{D t}{ab}\right)^{1/4} \,e^{-2^{3/2} \sqrt{\frac{D t}{ab}}}\,,\quad t\rightarrow \infty\,.\label{eq:saddleMe}
 \end{align}
To obtain the average length of the convex hull, we need to use the anisotropic Cauchy's formula (\ref{eq:LC}) as the maximum of Brownian motion in an ellipse is no more isotropic. However, as the average fluctuations (\ref{eq:saddleMe}) do not depend on the direction, we find
\begin{align}
  \langle L(t)\rangle &= \int_0^{2\pi} d\theta \,\langle M(\theta,t)\rangle \nonumber \\
  &\sim  \int_0^{2\pi} d\theta \,\xi(\theta)[1-\langle  \Delta(t)\rangle]\,,\quad t\to\infty\,,\label{eq:LE1}
\end{align}
where $\xi(\theta)$ is the maximal value of the maximum in the direction $\mathbf{e_\theta}$ for an ellipse. Performing the integral (\ref{eq:LE1}), we find
\begin{align}
  \mathcal{L}-\langle L(t )\rangle &\sim 2^{1/4}\sqrt{\pi } \, \mathcal{L}\,\mathcal{A}_2\,\left(\frac{D t}{ab}\right)^{1/4} \,e^{-2^{3/2} \sqrt{\frac{D t}{ab}}}\,,\quad t\to \infty \label{eq:Lell}\,,
\end{align}
where $\mathcal{L}$ is the perimeter of an ellipse given by
 \begin{align}
   \mathcal{L} = 4a \int_0^{\frac{\pi}{2}}d\theta\sqrt{1-e^2 \sin(\theta)^2}\,,\label{eq:Leff}
 \end{align}
 where $e=\sqrt{a^2-b^2}/a$ is the eccentricity of the ellipse. This time, we do not have a prediction of the exact expression for the amplitude. Nevertheless, the result is in good agreement with numerical data (see figure \ref{fig:decay}).

\section{Summary and outlook}\label{sec:ccl}
In this work, we studied the evolution of the maximum of a diffusive particle in confined environments in arbitrary dimensions. We first focused on the case of a particle confined in a $d$-dimensional ball of radius $R$. By relying on results on the NET, we showed that the behavior of the fluctuations of the maximum for $t\to \infty$ and close to $R$ exhibits a rich variety of behaviors depending on the dimension $d$. We then focused on the particular case of $d=2$ and applied our results to study the growth of the convex hull of Brownian motion in a disk with reflecting boundaries. Interestingly, we showed that it converges slowly to $2\pi R$ with a stretched exponential behavior. Finally, we discussed generalisations of our results to more general domains, such as the ellipse in two dimensions. 

It would be interesting to investigate further the extreme value statistics of Brownian motion in confined geometries. For instance, one could study the effect of confinement on the growth of the area of the convex hull \cite{Cauchy32,RandonFurling09,Majumdar10c}. Another possible extension of this work would be to study the record statistics of Brownian motion confined in a bounded domain. Finally, it would be interesting to extend our results to the case of Lévy flights and study how the fluctuations of the maximum are affected.
\section*{Acknowledgments}
We thank Bob Ziff for many inspiring discussions
over the years on various topics of random walks and Brownian motion. This work was partially supported by the Luxembourg National Research Fund (FNR) (App. ID 14548297).

\appendix 
\section{Numerical check of the  narrow escape time distribution}
\label{app:check}
\begin{figure}[h]
  \begin{center}
    \includegraphics[width=0.4\textwidth]{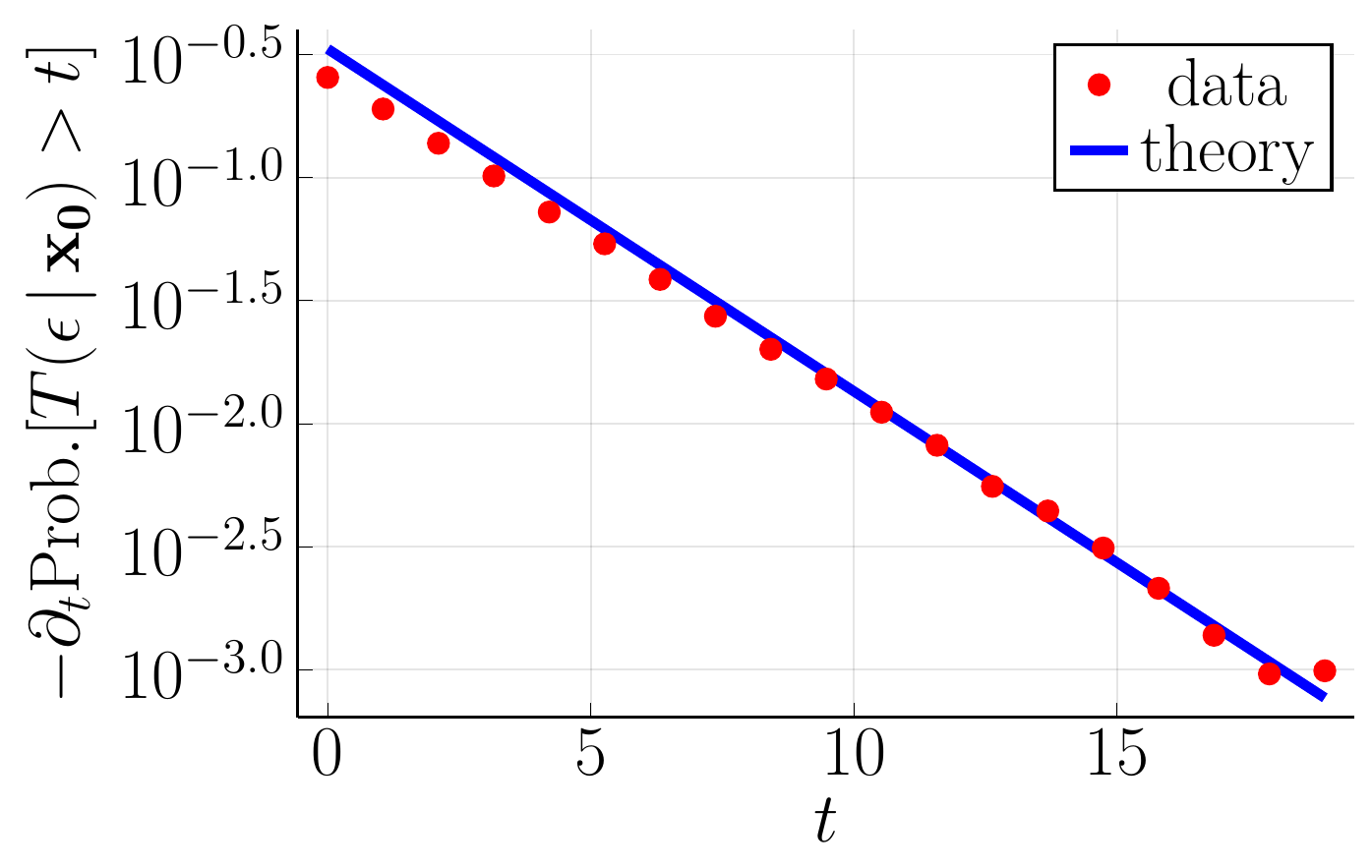}
    \caption{Numerical check of  the distribution of the narrow escape time (\ref{eq:PepsT}) for a Brownian motion in a disk of radius $R$ with a small opening of angle $2\epsilon$ on the boundary as depicted in figure \ref{fig:circle} for $\epsilon=0.1$ with $D=1$, $R=1$ and averaged over $10^5$ realisations with $dt=10^{-5}$.}
    \label{fig:check}
  \end{center}
\end{figure}

In this appendix, we check numerically the distribution of the NET (\ref{eq:PepsT}) for a Brownian motion in a disk with a small opening of angle $2\epsilon$ on the boundary as depicted in figure \ref{fig:circle}. The numerical data is in good agreement with the distribution (see figure \ref{fig:check}).

\section{Average fluctuations of the maximum in a disk}\label{app:saddle}
In this appendix, we compute the integral (\ref{eq:avgDelta2}) by the saddle point method. Let us first perform a change of integration variable $\chi_2=Ru/\sqrt{2Dt}$ which gives
\begin{align}
   \langle \Delta(t )\rangle &=  A_2\,\sqrt{\frac{2Dt}{R^2}}\,\int_0^\infty \frac{du}{u^2}\,\exp\left[-\sqrt{\frac{2Dt}{R^2}}\left( u+\frac{1}{u}\right)\right]\,.\label{eq:avgDelta2a}
\end{align}
Let us denote the function inside the exponential by
\begin{align}
  f(u) = u + \frac{1}{u}\,,\label{eq:fu}
\end{align}
which has a minimum at $u=1$ and is locally approximated by
\begin{align}
  f(u) \sim 2 + \frac{1}{2}\,(u-1)^2\,,\quad u\to 1\,.\label{eq:fu1}
\end{align}
Inserting the expansion (\ref{eq:fu1}) in (\ref{eq:avgDelta2a}) gives
\begin{align}
   \langle \Delta(t )\rangle &\sim  A_2\,\sqrt{\frac{2Dt}{R^2}}e^{-2\sqrt{\frac{2Dt}{R^2}}}\,\int_0^\infty \frac{du}{u^2}\,\exp\left[-\frac{\sqrt{2Dt}}{R}\,\frac{1}{2}(u-1)^2\right]\,,\quad t\to \infty\,.\label{eq:avgDelta2a2}
\end{align}
Finally, performing a change of variable $v=(2Dt/R^2)^{\frac{1}{4}}(u-1)$ and letting $t\to\infty$, we obtain
\begin{align}
   \langle \Delta(t )\rangle &\sim A_2\,\left(\frac{2Dt}{R^2}\right)^{\frac{1}{4}}\,e^{-2\sqrt{\frac{2Dt}{R^2}}}\,\int_{-\infty}^\infty dv\, e^{-\frac{v^2}{2}}\,,\quad t\to\infty\,,\label{eq:avgDelta2a3}
\end{align}
which, upon performing the Gaussian integral, recovers the expression (\ref{eq:saddleMxi}) displayed in the introduction.

\end{document}